# Raman Spectroscopy of $Bi_2Se_{3-x}Te_x$ (x= 0-3) Topological Insulator Crystals

Deepak Sharma[1, 2], M.M. Sharma[1, 2], R.S. Meena[1, 2], and V.P.S. Awana[1, 2, #]

[1]*CSIR-National Physical Laboratory, Dr. K.S. Krishnan Marg, New Delhi-110012, India*
[2]*Academy of Scientific and Innovative Research (AcSIR), Ghaziabad-201002, India*

**Abstract**

We report crystal growth and Raman spectroscopy characterization of pure and mixed bulk topological insulators $Bi_2Se_{3-x}Te_x$ (x= 0-3). The series comprises of both binary ($Bi_2Se_3$, $Bi_2Te_3$) and ternary tetradymite ($Bi_2Se_1Te_2$, $Bi_2Se_2Te_1$, $Bi_2Se_{1.5}Te_{1.5}$) topological insulators. We analyzed in detail the Raman peaks of vibrational modes viz. out of plane $A_1^1g$, $A_1^2g$, and in-plane $E_g^2$ for both binary and ternary tetradymite topological insulators. Both out of plane $A_1^1g$ and $A_1^2g$ exhibit obvious atomic size-dependent peak shifts and the effect is much lesser for the former than the latter. The situation is rather interesting for in-plane $E_g^2$, which not only shows the shift but rather a broader hump-like structure. The de-convolution of the same show two clear peaks, which are understood in terms of the presence of separate in-plane Bi–Se(I) and Bi-Te(I) modes in mixed tetradymite ($Bi_2Se_2Te_1$ and $Bi_2Se_{1.5}Te_{1.5}$) topological insulators. Summarily, various Raman modes of well-characterized pure and mixed topological insulator single crystals are reported and discussed in this article.

[#] Corresponding Author

Dr. V.P.S. Awana; E-mail- awana@nplindia.org
Ph. +91-11-45609357, Fax +91-11-45609310
Home page: awanavps.webs.om

## 1 Introduction

Ever since the discovery of bulk 3-D topological insulators (TIs), they remain the hot topic in condensed matter physics. The feature that TIs have time-reversal symmetry (TRS) protected conducting surface states with insulating bulk, provides immense opportunity to explore the new quantum phenomenon. TRS is very vulnerable to the external magnetic field and it is protected by intrinsic spin-orbit coupling (SOC) present in 3-D TIs [1-4]. These surface states are odd in number and the outermost state is termed as massless Dirac Fermion state [5].

These surface states can be seen in form of upper and lower halves of a cone in ARPES spectra, these cones are known as Dirac cones and the point where these cones meet is known as Dirac point [6]. These surface states discard any possibility of backscattering due to the presence of non-magnetic impurity. The presence of insulating bulk in 3-D TIs causes the dissipationless flow of current through these surface states, this property of 3-D TIs makes them very valuable in the spintronics arena [7].

In regards to 3-D TIs, some binary tetradymite compounds such as $Bi_2Se_3$, $Bi_2Te_3$, and $Sb_2Te_3$ emerged as model materials [8] and are highly studied in recent years. The bulk of these binary tetradymite TIs is not fully insulating but semiconducting in nature and it has been seen that their bulk insulating property can be enhanced by converting them into mixed TIs or ternary tetradymite materials such as $Bi_2Te_2Se$ and $Bi_2TeSe_2$ [9,10]. These ternary tetradymite materials have the crystal structure similar to binary tetradymite materials with quintuple layers separated by weak vander waals gap. The key feature in the structure of these ternary tetradymite materials is the position of Se atom, the middle Se atom layer remains intact in these ternary tetradymitematerials and is responsible for their enhanced insulating bulk properties [11]. These materials due to their insulating bulk have great importance in the study of surface conductance of TIs. Despite the importance of mixed TIs, there are not many reports based on the randomness of outer Se and Te layers in the unit cell, this randomness in local crystal structure affect crystal symmetries. These crystal symmetries have a direct impact on topological properties of TIs [12]. Raman Spectroscopy is a handy technique to study vibrational modes of crystals, which are sensitive to local crystal symmetries. Therefore, it is important to study the Raman Spectra of mixed TIs to get an insight intothe local crystal structure of these materials.

Keeping in view, the importance of mixed topological insulators in terms of an insight into their local crystal structure, we report here the crystal growth and Raman spectroscopy of ternary tetradymite TIs; $Bi_2Te_2Se$, $Bi_2TeSe_2$, $Bi_2Te_{1.5}Se_{1.5}$ and compare the same with that of binary tetradymite $Bi_2Se_3$ and $Bi_2Te_3$. Further, the growth parameters and structural details of all the studied compounds are given and discussed in this article. It is found that apart from the atomic size-dependent shift of various Raman modes, a broad hump is observed at the position of $E_g^2$ mode resulting in two clear peaks originating independently from Bi-Se(I) and Bi-Te(I).

## 2 Experimental

The solid-state reaction route is followed to grow single crystals of pure and mixed TIs. High purity (4N) powders were taken according to their stoichiometric ratio and grounded thoroughly in Argon gas filled MBRAUN glove box. Then these grounded samples were palletized with the help of hydraulic palletizer and vacuum-sealed in a quartz ampoule at a pressure of $5*10^{-4}$ mbar. Two different optimized heat treatments were followed to grow pure and mixed TIs samples [13]. Pure $Bi_2Se_3$ and $Bi_2Te_3$ samples were heated up to $950^0C$ at a rate of $120^0C/h$ and kept at this elevated temperature for 24 hours. This homogenous melt was slowly cooled at a rate of $2^0C/h$ to $650^0C$ and then allowed to cool normally to room temperature. The schematic of heat treatment for pure TIs is shown in Fig. 1(a). The heat treatment followed to grow mixed TI sample differs from above, besides hold time at $950^0$C, an extra hold period of 24 hours at $650^0$C was introduced. These samples were kept at $950^0$C for 48 hours and cooled to $650^0$C at a rate of $1^0$C/h. After a hold of 24 hours at $650^0$C; these samples were furnace cooled to room temperature. This extra hold time at $950^0$C and $650^0$C allows the melt to be homogenous as there is a mixing of elements in ternary tetradymite compounds and a relatively slow cooling compared to binary tetradymite compounds gives ample time for extra added elements to occupy desired positions so that these elements do not remain unreacted. The schematic of heat treatment followed for the growth of mixed TIs is shown in Fig. 1(b). Thus obtained single crystals are silvery shiny and easily cleavable along their growth direction. The image of the synthesized $Bi_2Te_3$ crystal is shown in Fig. 1(c).

Rigaku made Mini flex II X-ray diffractometer with $CuK_\alpha$ radiation of 1.5418Å wavelength is used to record X-ray Diffraction (XRD) pattern of both mechanically cleaved crystal flakes and gently crushed powder of synthesized crystals. Raman spectra of synthesized crystals are recorded by using Renishaw inVia Reflex Raman Microscope equipped with Laser of 514nm. & 720nm. The focal length of the spectrometer is 250mm and grating density is 2400 l/mm. The absolute accuracy of spectrometer is around $\pm 1 cm^{-1}$. Mechanically cleaved flakes of synthesized crystals are irradiated with a Laser of wavelength 514nm for 30 sec. During this; Laser power is maintained below 5 mW to avoid any local heating on the crystal surface. Powder XRD pattern is fitted by using Full Proof Software and VESTA software is used to draw unit cell structures of synthesized crystals.

## 3 Results

XRD pattern of mechanically cleaved crystal flakes of $Bi_2Se_3$, $Bi_2Te_3$, and $Bi_2Se_2Te_1$ are shown in Fig. 2. These XRD patterns confirm the uni-directional growth of as-grown crystals as high-intensity sharp peaks are observed only for (003n) plane, this shows that the crystals are grown in c-direction with high crystallinity. It is visible in Fig.2 that XRD peaks are shifted to the lower angle side as Te is substituted at Se site in $Bi_2Se_3$ lattice due to the bigger size of Te atom in comparison to Se atom. To maintain the clarity of the Figure, not all studied crystal flakes XRD are shown. Cleaved flakes synthesized crystals are not perfectly flat due to this some small unindexed peaks of planes other than (003n) planes are also observed in XRD pattern the presence of planes other than (003n) directions. It is worth mentioning that reflections from (009) and (0012) arenot observed for $Bi_2Te_3$ crystal flake.

Rietveld refinement of PXRD patterns of all studied single crystalline samples viz. $Bi_2Se_3$, $Bi_2Se_2Te$, $Bi_2Se_{1.5}Te_{1.5}$, $Bi_2SeTe_2$, and $Bi_2Te_3$ is performed by using Full Proof software and the results are shown in Fig. 3. Rietveld refinement confirms that the synthesized crystals are crystallized in the rhombohedral structure having R-3m space group. The quality of fit is estimated by $\chi^2$ (parameter for the goodness of fit), which is estimated to be 2.49, 1.58, 5.62, 2.81, and 3.27 for $Bi_2Se_3$, $Bi_2Se_2Te$, $Bi_2Se_{1.5}Te_{1.5}$, $Bi_2SeTe_2$, and $Bi_2Te_3$ respectively. These values are in an acceptable range and show the quality of crystals. All peaks in these PXRD patterns are indexed with their respective planes. Lattice parameters and atomic positions obtained from Rietveld refinement are listed in Table 1. It can be seen that lattice parameters are slightly increased as Te substitutes Se in $Bi_2Se_3$ unit cell because Te atom has a bigger size as compared to Se atom and hence this substitution increases the size of the unit cell.

Unit cells of as-grown $Bi_2Se_3$, $Bi_2Se_2Te$, $Bi_2Se_{1.5}Te_{1.5}$, $Bi_2SeTe_2$, and $Bi_2Te_3$ were drawn by using VESTA software and are shown in Fig. 4(a), 4(b), 4(c), 4(d) and 4(e) respectively. All these unit cells contain quintuple layers having Bi and Se/Te alternating atomic layers. Quintuple layers in these compounds can be represented as $-A_{VI}^{(I)} - B_V - A_{VI}^{(II)} - B_V - A_{VI}^{(I)} -$ where ($A_{VI}$ = Se, Te), and ($B_V$ = Bi) [11]. Subscripts of A and B represent the group of elements and $A_{VI}^{(II)}$ acts as the inversion center. Two adjoining quintuple layers are the mirror image of each other and separated by the vander waals gap. This gap provides the opportunity to intercalate a suitable element to get desired properties such as superconductivity [14, 15], etc. It can be seen in unit cells of mixed TIs that the middle Se layer remains intact in all cases this is due to the high electronegativity of Se. The high electronegativity of selenium as compared to tellurium makes Se more ionic in nature and its tendency to attract more electrons from $Bi^{3+}$ makes Se(II) position more preferable for Se as it is bonded with two of Bi atoms. Also the bond length of $B_V - A_{VI}^{(II)}$ is smaller than that for $B_V - A_{VI}^{(I)}$ [16] which shows that $B_V - A_{VI}^{(II)}$ bond is more ionic in nature which requiresa greater difference of electronegativity between two atoms and Se being more electronegative than Te fulfills this requirement and preferably occupy $A_{VI}^{(II)}$ in QL. This middle Se layer helps to increase bulk insulating property and the outermost Te or Se layers are responsible for the surface conduction [11]. The clear atomic positions of Bi, Se(I), Se(II), and Te(I) and Te(II) for both pure and mixed single-crystalline TIs are depicted in Fig. 4(a-e).

Fig. 5(a-d) shows Raman active vibrations of $Bi_2Se_3$, $Bi_2Te_3$, $Bi_2Se_2Te$ and $Bi_2Se_{1.5}Te_{1.5}$. Atomic vibrations in these TIs lead to four Raman active modes viz. $E_g^1$, $E_g^2$, $A_{1g}^1$, and $A_{1g}^2$. Out of plane stretching of atoms in the outer atomic layer i.e. ($A_{VI}^{(I)}$-$B_V$) or ($B_V$-$A_{VI}^{(I)}$) atoms leads to $A_{1g}$ mode. $E_g$ Raman active modes arise due to symmetric in-plane bending and shearing of atoms in the outer two layers i.e. ($A_{VI}^{(I)}$-$B_V$) or ($B_V$-$A_{VI}^{(I)}$). In Raman active $A_{1g}$ mode, atoms vibrate in opposite directions as resulting in a short displacement of atoms and produces higher phonon frequency. In Raman active $E_g$ mode, atoms vibrate in the same directions and results in greater atomic displacement of atoms producing lower phonon frequency [17, 18]. Interestingly, during these atomic vibrations, the middle atomic layer of Se (II) remains at rest.

Fig. 6 shows recorded Raman spectra of synthesized crystals of $Bi_2Se_3$, $Bi_2Se_2Te$, $Bi_2Se_{1.5}Te_{1.5}$, and $Bi_2Te_3$. Group theory allows four Raman active modes to be observed in bulk $Bi_2Se_3$ and $Bi_2Te_3$ crystals and these are $E_g^1$, $A_{1g}^1$, $E_g^2$, and $A_{1g}^2$ modes [18, 19]. Among these, the $E_g^1$, mode, which occurs below 50cm$^{-1}$ [20] could not be detected in our measurements due to instrumental limitations. Except for $E_g^1$, all other modes are observed in our measurements at their respective positions. Raman active $A_{1g}^1$, $E_g^2$, and $A_{1g}^2$ modes for synthesized $Bi_2Se_3$ crystals are observed at 72.5, 131.5,172.5cm$^{-1}$ and for $Bi_2Te_3$ crystal, the same is observed at 61.5, 101.5 and 133.5cm$^{-1}$ respectively; these results are in agreement with earlier reports [21]. In the case of mixed TI ($Bi_2Se_2Te$), the $A_{1g}^1$ and $A_{1g}^2$ modes are obtained at 68.74 and 165.04cm$^{-1}$ while these are observed at 67 and 160.6cm$^{-1}$ for $Bi_2Se_{1.5}Te_{1.5}$ and these are also in agreement with earlier reported results [18]. Interestingly, at the position of $E_g$ mode, a broader hump-like structure is observed for $Bi_2Se_2Te$ and $Bi_2Se_{1.5}Te_{1.5}$ in a range of 113-130cm$^{-1}$ and 110-128cm$^{-1}$ respectively. These humps are encircled in Fig. 6 and are explained in the discussion part. As it is clear from Fig. 6 Raman peaks in mixed TIs viz. $Bi_2Se_2Te$ and $Bi_2Se_{1.5}Te_{1.5}$ are shifted towards lower frequency as we substitute more Te at Se site in $Bi_2Se_3$. These shifts are marked by straight lines in Fig. 6. All observed Raman modes for pure and mixed synthesized crystals are listed in Table 2.

The humps that are observed at the position of $E_g^2$ mode for $Bi_2Se_2Te$ and $Bi_2Se_{1.5}Te_{1.5}$ suggest that $E_g^2$ mode in mixed TIs i.e. $Bi_2Se_{3-x}Te_x$ depends on stoichiometry. These humps are de-convoluted by using the Gausspeak fitting formula and are shown in Fig. 7(a) & 7(b). The de-convoluted plots clearly show the presence of two peaks at the position of $E_g^2$ mode. For peak1, and peak2, the full width at half maximum (FWHM) valuesare 8.72987cm$^{-1}$, 26.83004cm$^{-1}$, and 10.38092cm$^{-1}$, 48.41516cm$^{-1}$ for $Bi_2Se_2Te$, and $Bi_2Se_{1.5}Te_{1.5}$ respectively.Two peaks at the position of $E_g^2$ mode for mixed topological insulators are reported earlier as well [18].

## 4 Discussion

The results can be summarized in brief as follows;

1. The pure and mixed bulk topological insulators $Bi_2Se_{3-x}Te_x$ (x= 0, 1, 1.5, 2, 3) can be grown by solid-state reaction self-flux growth method with slightly different heat treatment for mixed TIs, than the pure ones i.e., x =0 and 3.
2. All three characteristic Raman modes namely out of plane $A_1^1g$, $A_1^2g$, and in-plane $E_g^2$ are seen for both pure and mixed TIs and these get shifted to lower frequency side with an increase in Te content. The shift is largest for $A_1^2g$ followed by $E_g^2$ and $A_1^1g$.
3. Although both $A_1^1g$ and $A_1^2g$ Raman modes are seen as distinct ones, the $E_g^2$ appeared as a broader hump for mixed TIs.

These results can be understood with the help of Fig. 5 (a-d), whereby different Raman active modes for the studied $Bi_2Se_{3-x}Te_x$ (x= 0, 1, 1.5, 2, 3) are shown. Fig. 5(a) and 5(b) depicts the out of plane $A_1^1g$ and $A_1^2g$ and in-plane $E_g^1$ and $E_g^2$ modes for $Bi_2Se_3$ and $Bi_2Te_3$ being represented by Bi-Se(I) and Bi-Te(I). Three out of four allowed Raman active modes viz. $E_g^2$, $A_{1g}^1$, and $A_{1g}^2$ are observed and are shown in Fig.6. It is clear from Fig. 6 that doping of Te atoms on Se site shifts the Raman peaks to lower frequencies. This can be explained as, that when Te is doped on Se site, Te atoms start to replace Se(I) atoms and results in Bi-Te(I) bond formation. Bonding force in Bi-Te(I) is weaker than that in Bi-Se(I) because Te atoms are less electronegative and larger than Se atoms. This change in bonding forces results in a shift in frequency of Raman modes to lower frequencies for mixed TIs i.e. $Bi_2Se_{3-x}Te_x$. As Te concentration increases, Raman modes shift more towards lower frequencies. In the case of $Bi_2Te_3$, Te atoms replace both Se(II) and Se(I) atoms resulting in the maximum shift in this case.

Further, it is seen from Fig. 6 that the shift in $A^1_{1g}$ mode is least than that in $E_g^2$ and $A^2_{1g}$ modes, the reason is vested in the origin of these modes. It is clear from fig. 5(a-d) that in the out of plane $A^1_{1g}$ mode, the bonds between outer two layers i.e. Bi-Se(I) (in case of $Bi_2Se_3$) vibrate in the same direction so polarizability (ease of distorting electrons from their original positions) of Bi-Se(I) does not change much during vibrations. This is the reason that frequency of $A^1_{1g}$ mode mainly depends on Bi-Se(II) bonding forces [18-21] because middle Se(II) layer remains at rest when Bi atom vibrates in the phase of Se(I) atom, which changes polarizability of Bi-Se(II) bond appreciably. In case of mix TIs i.e. $Bi_2Se_{3-x}Te_x$ ( x = 1, 1.5, 2), the Te atoms are doped on outer Se site, the Te atoms replace Se(I) atoms in outer atomic layers, which do not have a large impact on $A^1_{1g}$ mode frequency, while middle Se(II) layer remains intact, thus resulting in a small frequency shift in $A^1_{1g}$ mode on Te doping. In the case of in-plane $E_g^2$ and out of plane $A^2_{1g}$ modes, the bonds between outer two atomic layers i.e. Bi-Se(I) (in case of $Bi_2Se_3$) as shown in Fig. 5(a-d) vibrate opposite to each other, this makes the role of the outer atomic layer very important as during vibrations the polarizability of Bi-Se(I) changes appreciably in comparison to Bi-Se(II) as Se(II) atomic layer is at rest. So any change in electronegativity and atomic size of atoms in the outer two atomic layers results in a greater shift in frequency of $E_g^2$ and $A^2_{1g}$ in comparison to $A^1_{1g}$. This is the reason that these high-frequency Raman modes shift more to the lower frequency when bonds between atoms in the outer two atomic layers changes from Bi-Se(I) to Bi-Te(I) on Te doping as the later has lesser bond strength than former. While an appreciable shift has been observed for lower frequency $A^1_{1g}$ mode of $Bi_2Te_3$ where Te atom substitute Se atom on Se(II) site.

The shape of $E_g^2$, which shows a broad hump like structures for mixed TIs i.e. for $Bi_2Se_2Te$ and $Bi_2Se_{1.5}Te_{1.5}$ encircled in Fig. 6. The De-convolution of these humps is shown in Fig. 7(a) and 7(b) for $Bi_2Se_2Te$ and $Bi_2Se_{1.5}Te_{1.5}$ respectively, these plots show the presence of two peaks which represent two-modebehavior of $E_g^2$ mode. The observed two peaks for $E_g^2$ mode for $Bi_2Se_2Te$ and $Bi_2Se_{1.5}Te_{1.5}$ differs in width as the first peak in narrower while the second peak is broader as observed from FWHM parameters of Table 2, the reason behind this clearly observed result is the mixing of atoms at site-(I). First peak of splitted $E_g$ mode arises due to vibration of Bi-Te(I) mode while the second one arise due to mixing of Se and Te atoms and its width is further increased as mixing is increased as observed for $Bi_2Se_{1.5}Te_{1.5}$.

It is also reported earlier in ref. 18 that in mixed TIs high-frequency modes viz. $E_g^2$ and $A^2_{1g}$ show two-mode behavior which depends on the stoichiometry of the compound. Also, an extra mode in $Bi_2Te_2Se$ mixed TI is reported in ref. 22 and 23, which they attributed as an unidentified local mode emerged due to impurity. In our case, significant splitting has occurred in $E_g^2$ mode, which is described here as decoupled $E_g^2$ modes of Bi-Te(I) and Bi-Se(I) bonds as both vibrate with different frequencies. As it is described earlier when Te is doped on Se site, it replaces Se in upper atomic layers which leads to two different layers in unit cell one of Se(I) and other Te(I) and results in the formation of two different Bi-Se(I) and Bi-Te(I) bonds. The first peak in hump can be attributed to the $E_g^2$ mode of Bi-Te(I) and the second peak is of $E_g^2$ mode of Bi-Se(I). This can be seen in fig. 6 that the second peak that occurred due to Bi-Se(I) vibrations in $Bi_2Se_2Te$ is significantly suppressed in $Bi_2Se_{1.5}Te_{1.5}$ as Te doping is increased. The first peak is observed to shift towards $E_g^2$ mode of $Bi_2Te_3$ with gradually suppressing the second peak of Bi-Se(I) bonds and it is shown to be absent for $Bi_2Te_2Se$ in ref. 22,23, where Te replaces both Se(I) layers, and only the middle Se(II) layer remains. In ref. 22 and 23, a sharp unsplitted peak for $E_g^2$ mode of $Bi_2Te_2Se$ can be clearly seen while splitting is observed for $A_{1g}^2$ mode for $Bi_2Te_2Se$ which also confirm the fact that these both higher frequency modes viz. $E_g^2$ and $A_{1g}^2$ show two mode behavior for mixed crystals due to decoupling of Bi-Se(I) and Bi-Te(I) modes. Another possibility of occurrence of these extra modes can be breaking of inversion symmetry due to the presence of anti-site defect and randomness of Se(I) and Te(I) layers as suggested in ref. 12. This symmetry breaking turns some IR active $A_u$ mode to be Raman active but these modes should be very weak if at all observed in bulk crystals, as the same strongly depends on the dimension of the crystal [19] and becomes more prominent when crystal dimension is reduced. It is important to note that the appearance of $A_u$ modes depends on the nano-crystallinity driven quantum confinement effects irrespective of pure or mixed crystals [12,19]. The presence of nano-crystallinity driven quantum confinement effects resulting in $Eg^2$ peak split in the present series of studied crystals is ruled out because the same does not appear for pure i.e., $Bi_2Te_3$ and $Bi_2Se_3$ crystals. Significant, peak split is evident in the de-convoluted plot of both mixed TIs which discard the possibility of these modes to be symmetry breaking IR modes. It is thus clear that the first peak in hump can be attributed to the $E_g^2$ mode of Bi-Te(I) and the second peak is of $E_g^2$ mode of Bi-Se(I). The appearance of two peaks is evident only when Se and Te are in competing ratio and nearly non-existent for smaller doping [18].

## 5 Conclusion

In this article, we made an analysis of the XRD pattern and Raman Spectra for binary tetradymite TIs viz. $Bi_2Se_3$ and $Bi_2Te_3$ and ternary tetradymite TIs of general formula $Bi_2Se_{3-x}Te_x$. Unidirectional growth of synthesized crystals is evident from the XRD pattern taken on mechanically cleaved crystal flakes. Raman spectra of both binary and ternary tetradymite TIs represented shifting in Raman modes due to doping of Te and broader hump-like structure representing the de-coupled $E_g^2$ modes of Bi-Te(I) and Bi-Se(I) bonds, while these humps are not observed in out of plane $A_1^1g$ and $A_1^2g$ modes.

**Acknowledgment**

The authors would like to thank Director CSIR-NPL for his keen interest and encouragement. Authors are also grateful to Mrs. Shaveta Sharma Sharda for timely Raman Spectroscopy, Mr. Naval Kishore Upadhyay for X-Ray Diffraction (XRD) measurements, and Mr. Krishna Mohan Kandpal for vacuum encapsulation. Deepak Sharma & M.M. Sharma would like to thanks CSIR for the research fellowship and AcSIR-Ghaziabad for Ph.D. registration.

Table 1. Rietveld refined lattice parameters and atomic positions of $Bi_2Se_3$, $Bi_2Se_2Te_1$, $Bi_2Se_{1.5}Te_{1.5}$, $Bi_2Se_1Te_2$, and $Bi_2Te_3$ single crystals.

| $Bi_2Se_3$ | $Bi_2Se_2Te_1$ | $Bi_2Se_{1.5}Te_{1.5}$ | $Bi_2Se_1Te_2$ | $Bi_2Te_3$ |
|---|---|---|---|---|
| a = 4.156(3) Å | a = 4.219(3) Å | a = 4.266(9) Å | a = 4.281(4) Å | a = 4.386(6) Å |
| b = 4.156(3) Å | b = 4.219(3) Å | b = 4.266(9) Å | b = 4.281(4) Å | b = 4.386(6) Å |
| c = 28.736(8) Å | c = 29.511(5) Å | c = 29.165(11) Å | c = 29.922(5) Å | c = 30.499(13) Å |
| $\alpha=\beta=90^0$ | $\alpha=\beta=90^0$ | $\alpha=\beta=90^0$ | $\alpha=\beta=90^0$ | $\alpha=\beta=90^0$ |
| $\gamma=120^0$ | $\gamma=120^0$ | $\gamma=120^0$ | $\gamma=120^0$ | $\gamma=120^0$ |
| Atom Positions | | | | |
| $Bi_2Se_3$ | $Bi_2Se_2Te_1$ | $Bi_2Se_{1.5}Te_{1.5}$ | $Bi_2Se_1Te_2$ | $Bi_2Te_3$ |
| Bi: 0,0, 0.3999 | Bi: 0,0,0.3978 | Bi: 0,0,0.4015 | Bi: 0,0, 0.3974 | Bi: 0,0,0.4038 |
| Se(I): 0,0,0 | Se(I):0,0,0.2120 | Se(I): 0, 0, 0.2020 | Se(II): 0,0,0 | Te(I): 0,0,0 |
| Se(II): 0,0,0.2117 | Se(II): 0,0,0 | Se(II): 0, 0, 0 | Te(I): 0,0,0.2123 | Te(II): 0,0,0.2039 |
| | Te(I): 0,0, 0.2120 | Te(I):0, 0, 0.2020 | | |

Table 2: Raman active modes of $Bi_2Se_3$, $Bi_2Se_2Te$, $Bi_2Se_{1.5}Te_{1.5}$, and $Bi_2Te_3$

| Raman peaks | | | | |
|---|---|---|---|---|
| Raman Modes | $Bi_2Se_3$ | $Bi_2Se_2Te$ | $Bi_2Se_{1.5}Te_{1.5}$ | $Bi_2Te_3$ |
| $A^1_{1g}$ | 73.0± 1cm$^{-1}$ | 69± 1cm$^{-1}$ | 67.0± 1cm$^{-1}$ | 62.0± 1cm$^{-1}$ |
| $E^2_g$ | 132.0± 1cm$^{-1}$ | 113± 1cm$^{-1}$ & 130 ± 1cm$^{-1}$ | 111.0± 1cm$^{-1}$ & 124.0± 1cm$^{-1}$ | 102.0± 1cm$^{-1}$ |
| $A^1_{2g}$ | 173.0± 1cm$^{-1}$ | 165.0± 1cm$^{-1}$ | 161.0± 1cm$^{-1}$ | 134.0± 1cm$^{-1}$ |
| FWHM Parameters | | | | |
| Raman Modes | $Bi_2Se_3$ | $Bi_2Se_2Te$ | $Bi_2Se_{1.5}Te_{1.5}$ | $Bi_2Te_3$ |
| $A^1_{1g}$ | 8.38 cm$^{-1}$ | 9.44 cm$^{-1}$ | 11.93432 cm$^{-1}$ | 5.72 cm$^{-1}$ |
| $E^2_g$ (Peak1) | 15.11 cm$^{-1}$ | 8.73 cm$^{-1}$ | 10.38 cm$^{-1}$ | 6.712 cm$^{-1}$ |
| $E^2_g$ (Peak2) | | 26.83 cm$^{-1}$ | 48.42 cm$^{-1}$ | |
| $A^1_{2g}$ | 19.65 cm$^{-1}$ | 17.57 cm$^{-1}$ | 28.19cm$^{-1}$ | 13.96 cm$^{-1}$ |

**Figure Captions**
1. (a) Schematic diagram of heat treatment to grow pure $Bi_2(Se/Te)_3$
   (b) Schematic diagram of heat treatment to grow $Bi_2Se_{3-x}Te_x$ (x=1, 1.5, 2) mixed TI crystals.
   (c) Image of as grown $Bi_2Te_3$ crystal.
2. X-ray diffraction pattern of mechanically cleaved flakes of $Bi_2Se_{3-x}Te_x$ (x = 0, 1 & 3) single crystals.
3. Rietveld refined Powder X-ray Diffraction (PXRD) pattern for $Bi_2Se_{3-x}Te_x$ (x = 0, 1, 1.5, 2 & 3) crystals.
4. Unit cell structure of (a) $Bi_2Se_3$ (b) $Bi_2Se_2Te$ (c) $Bi_2Se_{1.5}Te_{1.5}$ (d) $Bi_2SeTe_2$ and (e) $Bi_2Te_3$ single crystals.
5. Raman Active modes for (a) $Bi_2Se_3$ (b) $Bi_2Te_3$ (c) $Bi_2Se_2Te$ and (d) $Bi_2Se_{1.5}Te_{1.5}$.
6. (a) Raman Spectra of $Bi_2Se_3$, $Bi_2Se_2Te$, $Bi_2Se_{1.5}Te_{1.5}$, and $Bi_2Te_3$.
7. (a) De-convoluted Raman Active $E^2_g$ mode of $Bi_2Se_2Te$ mixed TI crystal.
   (b) De-convoluted Raman Active $E^2_g$ mode of $Bi_2Se_{1.5}Te_{1.5}$ mixed TI crystal.

# 6 References:

1. Y. Ando, J. Phys. Soc. Jap. **82**, 102001 (2013).
2. M. Z. Hasan and C. L. Kane, Rev. Mod. Phys. **82**, 3045 (2010).
3. T. Arakane,T. Sato, S. Souma, K. Kosaka, K. Nakayama, M. Komatsu, T. Takahashi, Zhi Ren, Kouji Segawa, Yoichi Ando, Nat. Comm. **3**, 636 (2012).
4. Stephan Rachel, Rep. Prog. Phys. **81**, 11 (2018).
5. Xiao-Liang Qi and Shou-Cheng Zhang, Rev. Mod. Phys. **83**, 1057 (2011).
6. D. Hsieh, Y. Xia, D. Qian, L. Wray, F. Meier, J. H. Dil, J. Osterwalder, L. Patthey, A. V. Fedorov, H. Lin, A. Bansil, D. Grauer, Y. S. Hor, R. J. Cava, and M. Z. Hasan, Phys. Rev. Lett. **103**, 146401 (2009).
7. Dmytro Pesin, Allan H. MacDonald, Nat. Mater. **11**, 409 (2012).
8. H. Zhang, C.-X. Liu, X.-L. Qi, X. Dai, Z. Fang, and S.-C. Zhang, Nature Phys. **5**, 438 (2009).
9. Z. Ren, A. A. Taskin, S. Sasaki, K. Segawa, and Y. Ando, Phys. Rev. B **82**, 241306 (2010).
10. Shu Cai, Jing Guo, Vladimir A. Sidorov, Yazhou Zhou, Honghong Wang, Gongchang Lin, Xiaodong Li, Yanchuan Li, Ke Yang, Aiguo Li, Qi Wu, Jiangping Hu, Satya. K. Kushwaha, Robert J. Cava & Liling Sun, npj Quant. Mater. **3**, 62 (2012).
11. R. J. Cava, Huiwen Ji, M. K. Fuccillo, Q. D. Gibsona, and Y. S. Horb J. Mater. Chem. C, **1**, 3176 (2013).
12. Raphael German, Evgenia V. Komleva, Philipp Stein, Vladimir G. Mazurenko, Zhiwei Wang, Sergey V. Streltsov, Yoichi Ando, and Paul H. M. van Loosdrecht, Phys. Rev. Mater. **3**, 054204 (2019).
13. R. Sultana, G. Awana, B. Pal, P. K. Maheshwari, M. Mishra, G. Gupta, A. Gupta, S. Thirupathaiah, and V. P. S. Awana, J. Sup. Nov. Mag. **30**, 2031 (2017).
14. M. M. Sharma, P. Rani, Lina Sang, X.L. Wang & V.P.S. Awana, J. Sup. Nov. Mag. **33**, 565 (2020).
15. M. M. Sharma, Lina Sang, Poonam Rani, X. L. Wang & V. P. S. Awana, J. Sup. Nov. Mag. **33**, 1243 (2020)
16. Seizo Nakajima, J. Phys. Chem. Solids **24**, 479 (1963).
17. J. Yuan, M. Zhao, W. Yu, Y. Lu, C. Chen, M. Xu, S. Li, and K. P. Loh, and Q. Bao Materials (Basel) **8**, 5007 (2015).
18. W. Richter, H. Kohler and C. R. Becker, Phys. Status Sol. (b) **84**, 619 (1977).
19. K. M. F. Shahil, M. Z. Hossain, V. Goyal, and A. A. Balandin, J. Appl. Phys. **111**, 054305 (2012).
20. V. Gnezdilov, Y. G. Pashkevich, H. Berger, E. Pomjakushina, K. Conder, and P. Lemmens, Phys. Rev. B **84**, 195118 (2011).
21. Rabia Sultana, Ganesh Gurjar, S Patnaik, VPS Awana, Mater. Res. Exp. **4**, 046107 (2018).
22. Yao Tian, Gavin B. Osterhoudt, Shuang Jia, R. J. Cava, Kenneth S. Burch, Appl. Phys. Lett. **108**, 041911 (2016).
23. Ana Akrap, Michael Tran, Alberto Ubaldini, Jeremie Teyssier, Enrico Giannini, and Dirk van der Marel, Phys. Rev. B **86**, 235207 (2012).

Fig. 1(a)

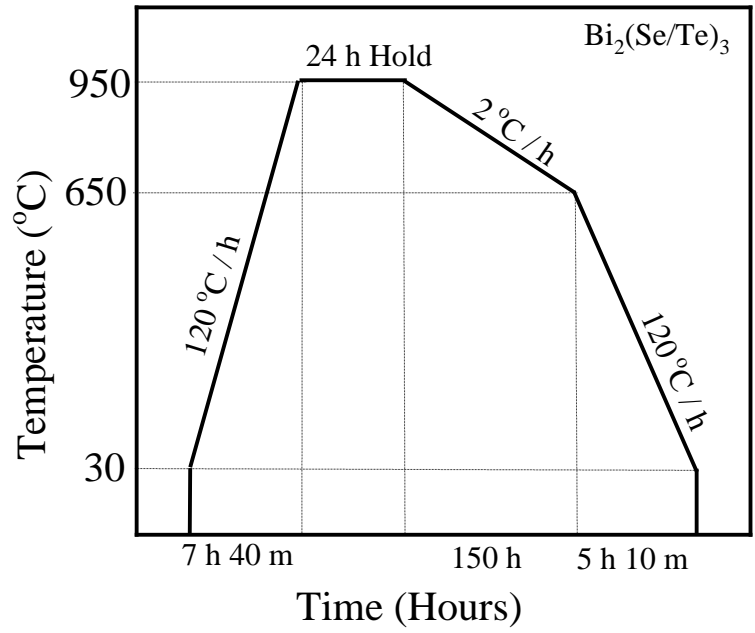

Fig. 1(b)

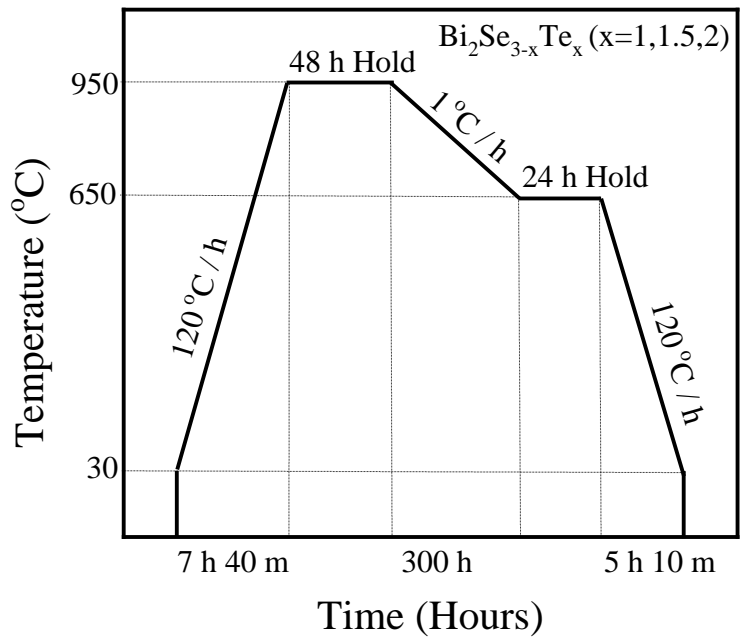

Fig. 1(c)

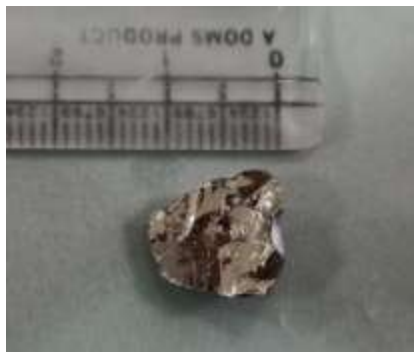

Fig. 2

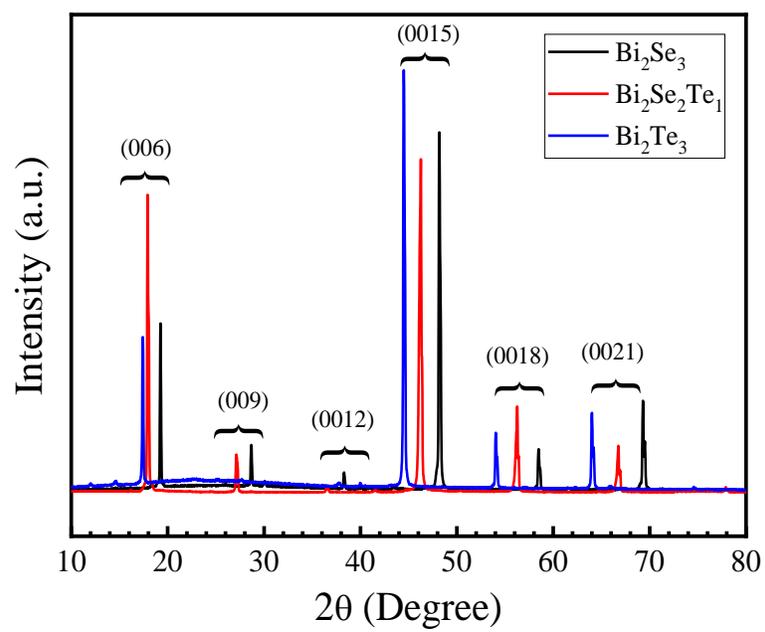

Fig. 3

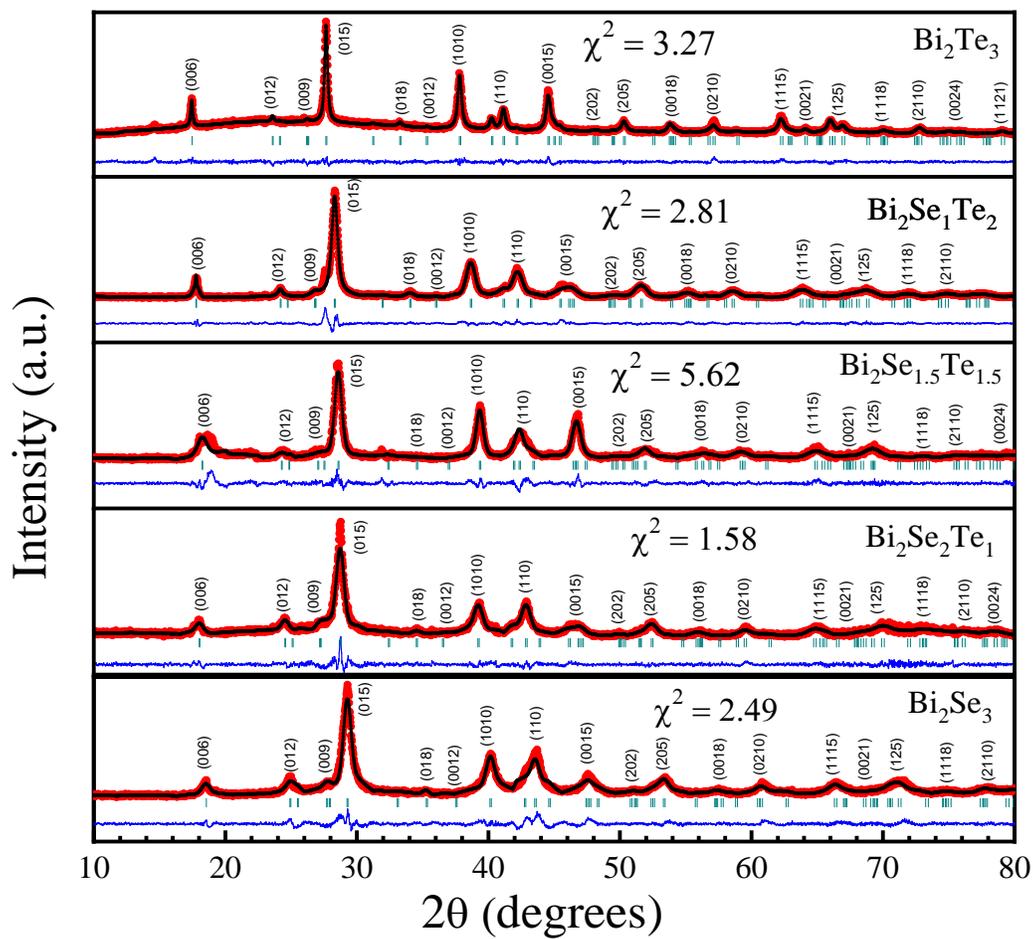

Fig. 4 (a)

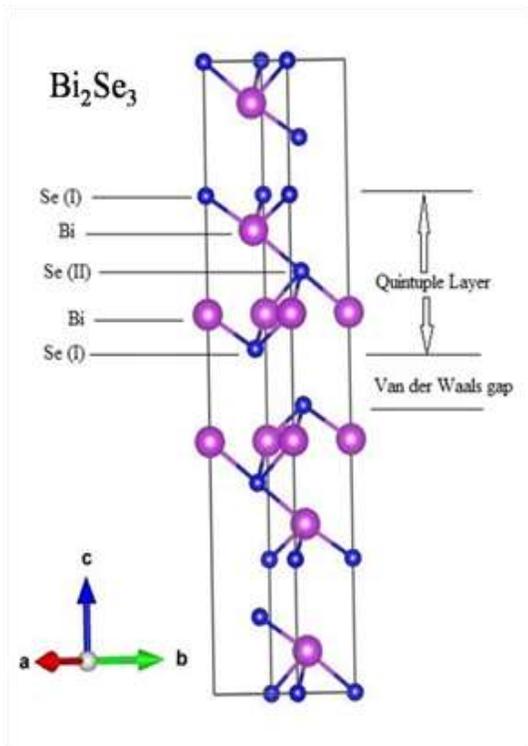

Fig. 4(b)

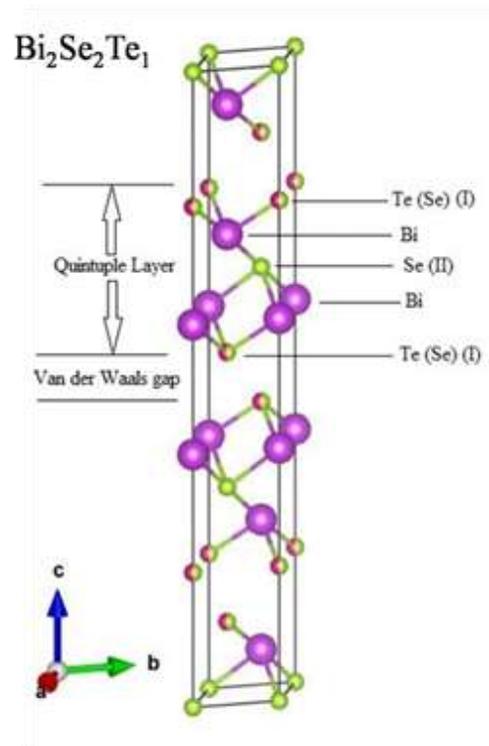

Fig. 4(c)

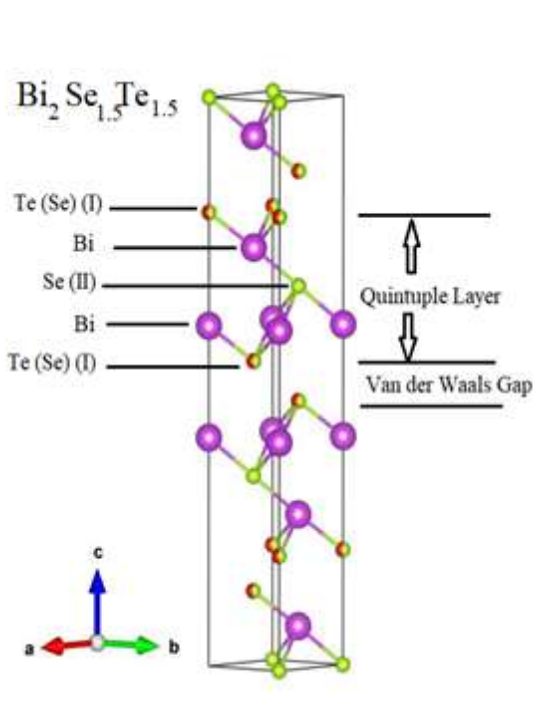

Fig. 4(d)

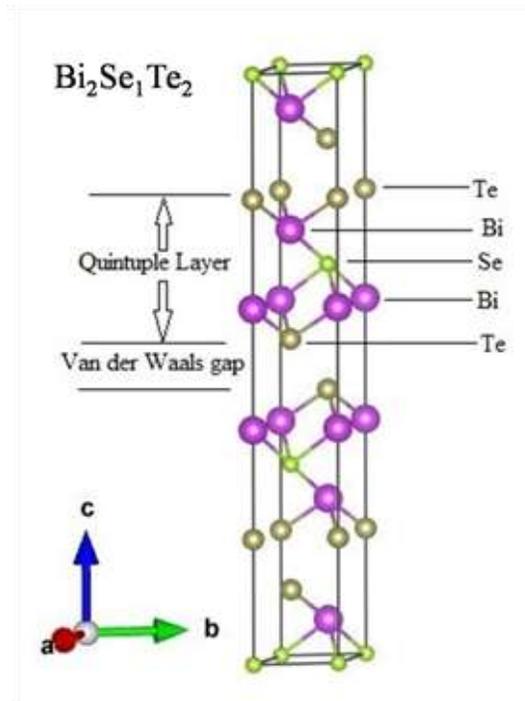

Fig. 4(e)

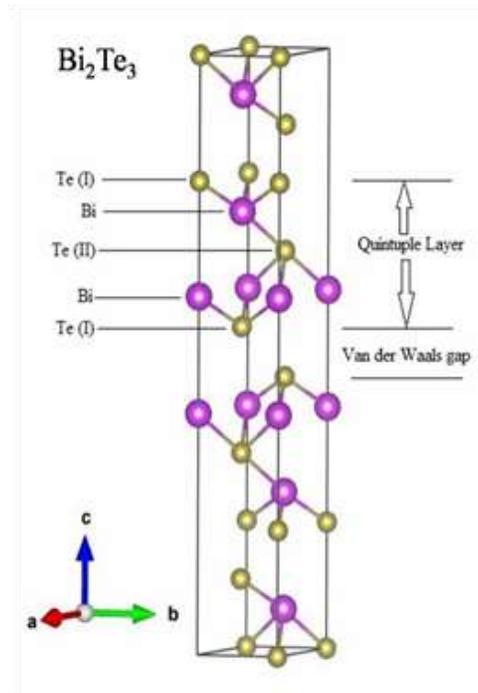

Fig. 5(a)

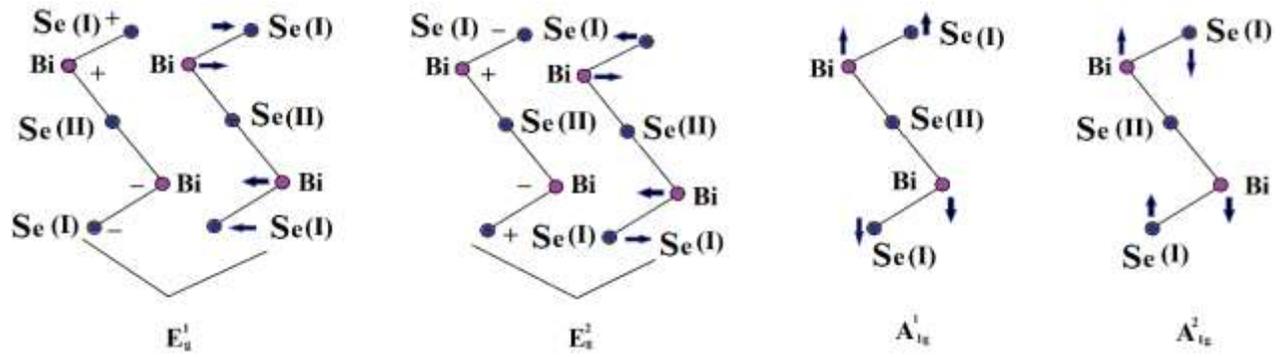

Fig. 5(b)

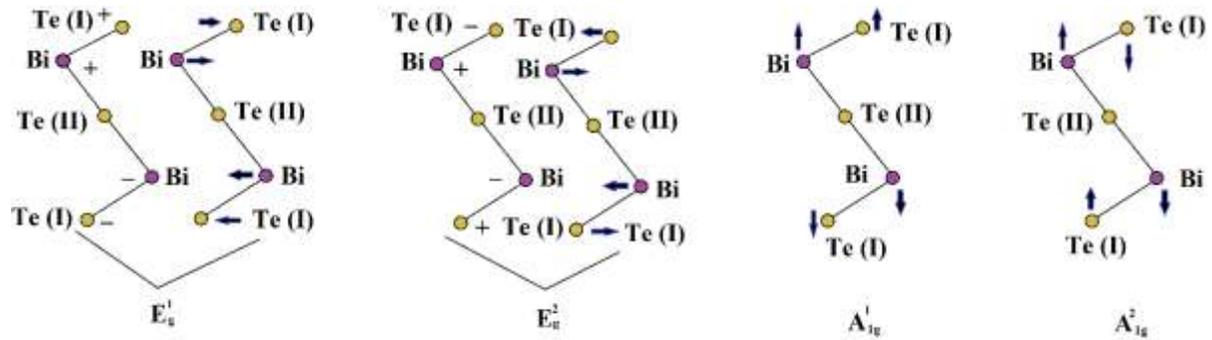

Fig. 5(c)

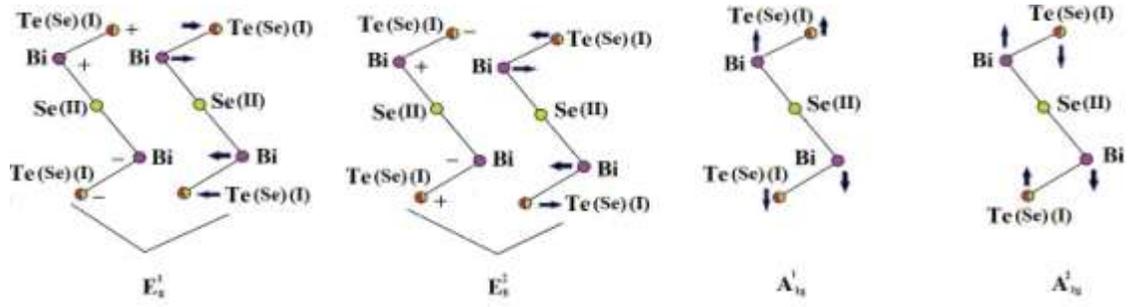

Fig. 5(d)

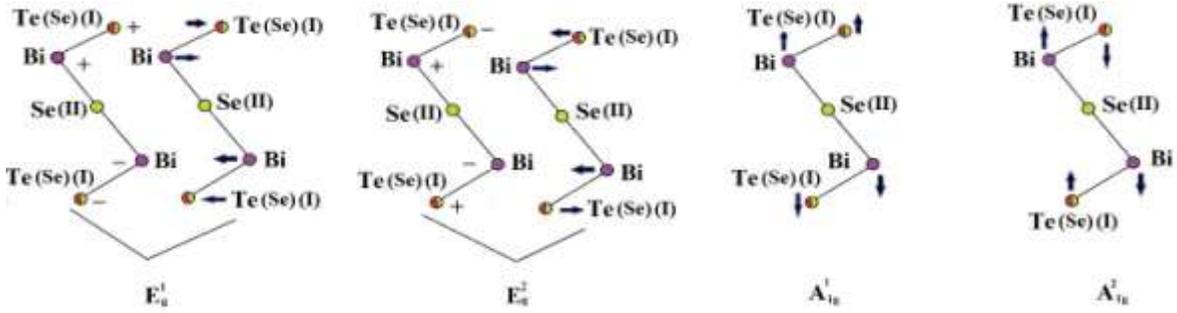

Fig. 6(a)

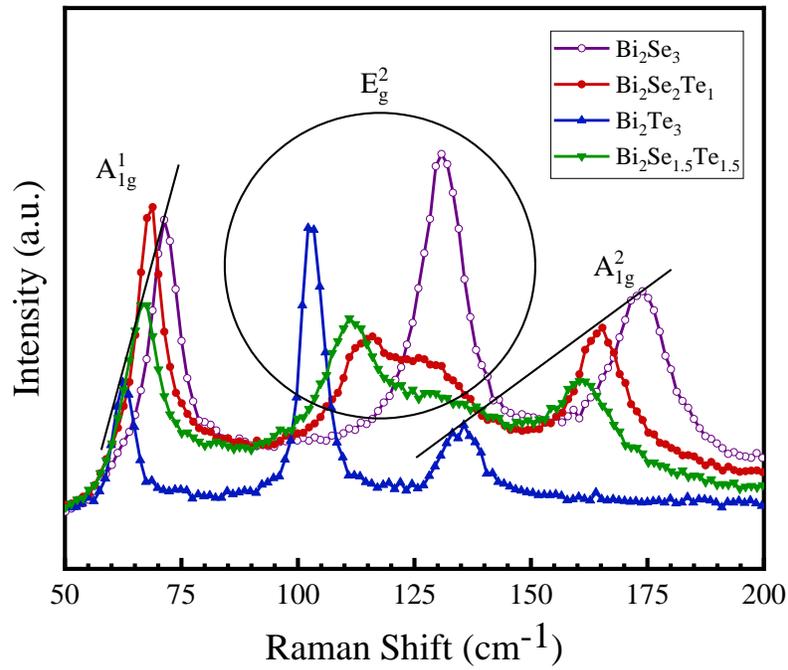

Fig. 7(a)

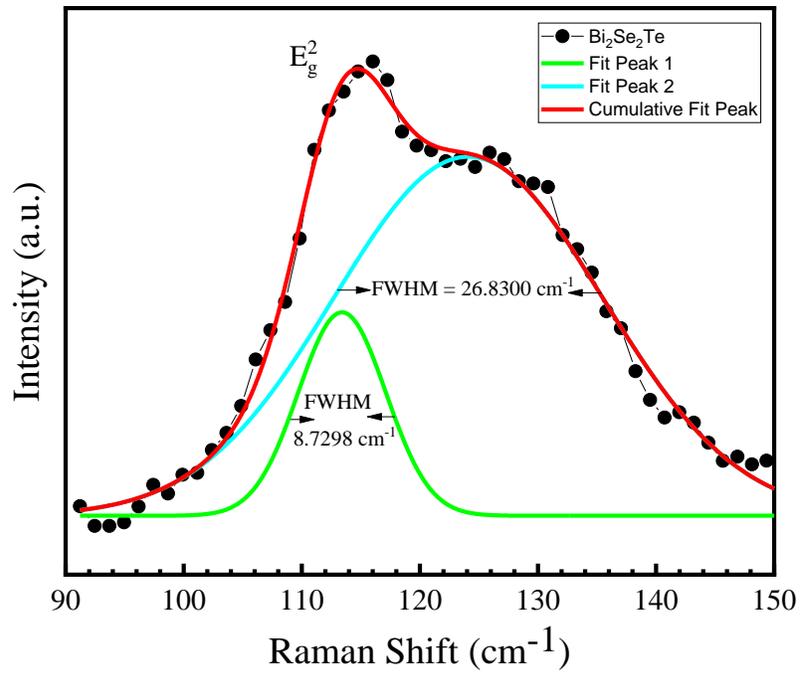

Fig. 7(b)

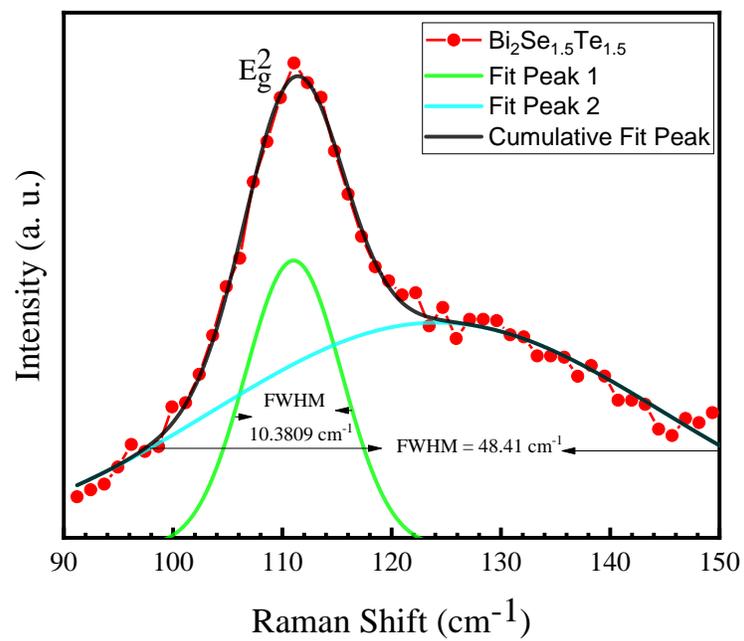